\documentclass[useAMS,usenatbib]{mn2e}
\usepackage{graphicx}
\title[CK\,Vul: nebula and three background stars]{CK\,Vul: evolving nebula and three curious background
stars}
\author[Hajduk et al.]{M.~Hajduk,$^1$
  P.A.M.~van Hoof,$^2$ A.A.~Zijlstra$^3$ \\
$^1$Nicolaus Copernicus Astronomical Center, 
ul. Rabia\'{n}ska 8, 87-100 Toru\'{n}, Poland \\
$^2$Royal Observatory of Belgium, Ringlaan 3, 1180 Brussels, Belgium \\
$^3$Jodrell Bank Centre for Astrophysics, Alan Turing Building,
Manchester M13 9PL, UK \\
}
\begin{document}

\date{}


%

\maketitle

\label{firstpage}

\begin{abstract} 







We analyse the remnants of CK\,Vul (Nova\,Vul\,1670) using optical imaging and
spectroscopy. The imaging, obtained between 1991 and 2010, spans 5.6\% of the
lifetime of the nebula. The flux of the nebula decreased during the last 2
decades. The central source still maintains the ionization of the innermost part
of the nebula, but recombination proceeds in more distant parts of the nebula.
Surprisingly, we discovered two stars located within $10^{\prime\prime}$ of the
expansion centre of the radio emission that are characterized by pronounced long
term variations and one star with high proper motion. The high proper motion
star is a foreground object, and the two variable stars are background objects.
The photometric variations of two variables are induced by a dusty cloud ejected
by CK\,Vul and passing through the line of sight to those stars. The cloud
leaves strong lithium absorption in the spectra of the stars. We discuss the
nature of the object in terms of recent observations.



\end{abstract}

\begin{keywords}
stars: AGB and post-AGB--binaries: general -- stars: evolution -- stars: individual: CK Vul -- stars: mass-loss -- planetary nebulae: general
\end{keywords}

\section{Introduction}

CK\,Vul erupted in 1670, becoming a 2.6 magnitude star at its maximum
\citep{Shara82}. With its two subsequent fadings and recoveries on the timescale
of one year, the 1670-72 lightcurve remains very intriguing as well as poorly
understood. The object was interpreted as a slow nova \citep{Shara85},
born-again star \citep{Harrison96,Evans02}, or finally, a stellar merger \citep{Kato03}. Recently, \citet{Miller11} proposed the diffusion-induced nova scenario for CK\,Vul.

The central star has not been observed since its decline in 1672
\citep{Naylor92}, leaving us with no chance to inspect the product of the
outburst directly. The current limit for the central object brightness is $r' >
23$\,mag \citep{Hajduk07}. The apparent amplitude of the outburst may have
possibly been as large as 20\,mag or more.

What we currently observe is a gaseous nebula with a maximum extent of the
symmetric, limb-brightened lobes of about $1^{\prime}$. The expansion
measurements have proven that it indeed dates back to the XVII century
\citep{Hajduk07}. Different parts of the nebula show enhanced [N{\sc ii}] lines
exceeding in intensity the H$\alpha$ emission. A faint, barely resolved radio
emission is detected at the centre of expansion.

We have discovered two variable stars located within $3 - 4$ arcsec of the
expansion centre in the plane of the sky. The apparent separation of the two
variable stars is about 2 arcsec. The coincidence of the position of the stars
with the expansion centre of the CK\,Vul nebula is striking. In addition, we
found a star with pronounced proper motion in the vicinity of the expansion
center of CK\,Vul. In this paper, we investigate the link of the variable and
high proper motion star with CKVul and discuss the nature of the object in the
light of new observations.

\section{Observations and data reduction}

CK\,Vul was observed with the WHT telescope with the Prime Focus Camera. The
service programme observations were carried in the R and H$\alpha$ 6559/57
filters on April 14th, 2009.

Imaging was also performed on the Gemini Multi-Object Spectrograph (GMOS) at
three different epochs in 2010 with the Sloan $g'r'i'z'$ wide band filters and
on one occasion with various emission lines and corresponding continuum filters
(Table \ref{images}). The transformation equations between the $UBVR_CI_C$ and
$g'r'i'z'$ systems were provided by \citet{Smith02}.

The data were reduced using the {\sc iraf gmos} package. The instrumental
magnitudes were derived using the {\sc iraf daophot} package. The absolute flux
calibration for the photometry was obtained using the 2010\,April\,23
observation of the SA\,110-361 Landolt standard field and the 2010\,June\,23
observation of the PG\,2213-006 field and the standard Mauna Kea extinction
curve. Six standard stars were present in the SA\,110-361 field and four in the
PG\,2213-006 field. The rms of the fit for the SA\,110-361 (PG\,2213-006) field
was 0.12 (0.06), 0.08 (0.05), 0.06 (0.02) and 0.06 (0.03) mag for the $g'$,
$r'$, $i'$ and $z'$ filters, respectively. 

The magnitudes for the remaining archive images (either in the Sloan or
Johnson-Morgan-Cousins system) were calibrated with respect to the
2010\,June\,23 and April\,23 observations using all the available field stars,
by means of a least squares linear fit.

The $r'$ and $i'$ magnitudes derived from the GMOS imaging were checked against
the IPHAS catalogue \citep{drew05}. For this purpose, IPHAS magnitudes were
converted from the Vega to the $AB_{95}$ system \citep{fukugita95}. The IPHAS
and GMOS absolute calibration agree to about 0.08 magnitude in $r'$ and 0.02
magnitude in $i'$.

Emission line images in the $\rm H\alpha$, [O{\sc\,iii}] and [S{\sc\,ii}] bands
were continuum subtracted using redshifted $\rm H\alpha$ (for $\rm H\alpha$ and
[S{\sc\,ii}]) and [O{\sc\,iii}] filters using the ISIS package, accounting for
different PSFs \citep{Alard98}.

\begin{table}
 \caption[]{A log of the CK\,Vul imaging observations. The data were taken with the 4.2-m William Herschel Telescope (WHT), the 2.54-m Isaac Newton Telescope (INT) and the 8.1-m Gemini North (GN).}
 \label{images}
 \begin{tabular}{@{}ccc}
  \hline
  date		&telescope	& filters used		\\
  \hline
  1991-08-10	&WHT		&{H$\alpha$} $R$			\\
  2004-09-04	&INT		&{H$\alpha$} $r'$			\\
  2005-07-12	&INT		&{H$\alpha$} $r' i'$		\\
  2009-04-14	&WHT		&{H$\alpha$} $R$			\\
  2010-04-23	&GN		&$g' r' i' z'$ 		\\
  2010-06-10	&GN		&$g' r' i' z'$ 		\\
  2010-06-22	&GN		&{H$\alpha$} [O{\sc iii}] [S{\sc ii}]\\
  		&		&redshifted {H$\alpha$} redshifted [O{\sc iii}]\\
  2010-06-23	&GN		&$g' r' i' z'$	 	\\
  \hline
 \end{tabular}
\end{table}

\begin{figure}
\includegraphics[width=84mm]{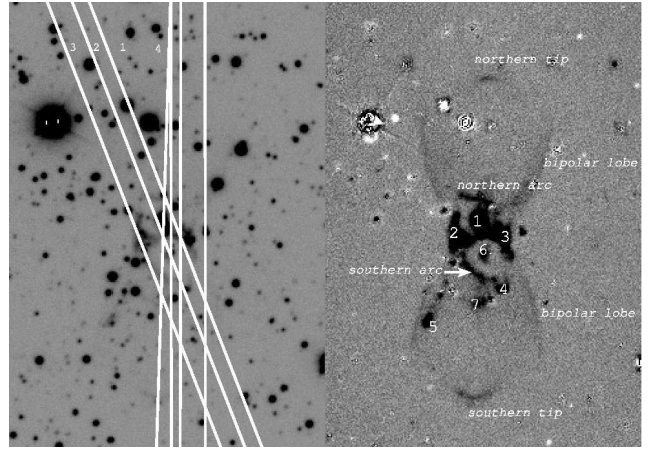}
\caption{\label{slits} The positions of the slit for the spectroscopic
observations on the background of the GMOS mosaic H$\alpha$ image (left). The slit width was wider than indicated in the figure. Four out of seven slit positions are numbered. The H$\alpha$ continuum subtracted image is shown on the right. The nebular components are named or numbered as in \citet{Hajduk07}. North is at the top, East to the left. The image scale is $100 \times 80$ arcsec.}
\end{figure}

Longslit spectra at seven different positions were taken with the GEMINI North
GMOS facility on June 10th, 12th and 14th, 2010. The slit positions are plotted
in Figure \ref{slits}. In four positions the slit was centered on the nebula,
including the three brightest nebulosities near the center and the tips of the
bipolar nebula. Three other observations were centered on the field stars. Only
the slit positions centered on the nebula are numbered in Figure~\ref{slits}.

The B600 grating was used and the central wavelength was set at 550 or 560\,nm.
The spectra were calibrated using a spectrophotometric standard star
BD+25\,3941  observed during spectroscopic nights. The mean extinction curve at
Mauna Kea was used. The reduction was performed using the {\sc iraf gmos}
package.

\begin{figure}
\includegraphics[width=84mm]{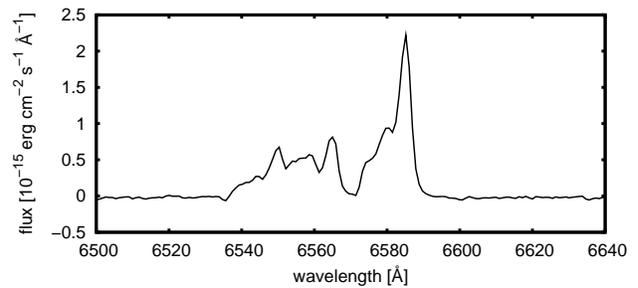}
\caption{\label{neb1} Spectrum of the southern part of blob~1 showing the $\rm H \alpha$ and [N\,{\sc ii}] lines.
}
\end{figure}


\section{The central star and the nebula of CK\,Vul}

\subsection{Morphology and kinematics of the nebula}

The central region of the nebula consists of a group of individual blobs and
arcs (Figure~\ref{slits}). Three brightest blobs (1, 2 and 3) are accompanied by
smaller blobs and arcs. All these components are embedded in a faint,
limb-brightened bipolar nebula resembling an hourglass. Two tips located on the
symmetric axis are the furthermost structures of the bipolar nebula.

The spectrum taken at the 1st position was centered on blob~3. Blob~3 can be
decomposed into two parts characterized by different radial velocities. The
southern part of the blob shows a radial velocity gradient along the slit. The
northern part of blob~3 is blue-shifted with respect to the southern part. Other
fainter parts of the nebula showing different velocities are observed north of
the brightest parts of blob~3. The slit cuts through the northern arc in this
position. The extended, faint emission seen only in the $\rm H \alpha$ and
[N\,{\sc ii}] lines originates in the western edge of the southern lobe.


The second slit includes blobs 1, 6, 4 and the northern arc. A blue-shifted,
separate velocity component is seen in the southern part of nebulosity~1. It is
indistinguishable from nebulosity~1 in the image (Figure \ref{neb1}).

The third slit cuts through the nebulosity~2, the southern arc and blob~7. The
faint edge of the northern lobe is seen above blob~2, showing a positive radial
velocity, increasing with the distance from the center of the nebula, and $\rm H
\alpha$ stronger than the [N\,{\sc ii}] 6584 \AA\ line. 

Nebulosity~2 is not a part of the bipolar nebula since it shows different
velocity gradient along the slit and negative (with respect to the systemic
velocity) velocities, while northern lobe of the bipolar nebula shows positive
velocities. Blob~2 is not limb-brightened, contrary to the bipolar nebula.

The fourth slit includes the tips of the bipolar nebula and blob~1. In the
southern part of blob~1, near the centre of the nebula, an additional, faint
component is seen at negative velocities. The same component is observed in the
second slit. The radial velocity of blob~1 decreases with the distance from the
center of the nebula. This suggests that blob~1 lies in the plane perpendicular
to the symmetric axis of the bipolar nebula.

The three remaining spectra, obtained at the PA of 0 degree, were centered on the two variable stars and the high proper motion star.

\begin{figure}
\includegraphics[width=84mm]{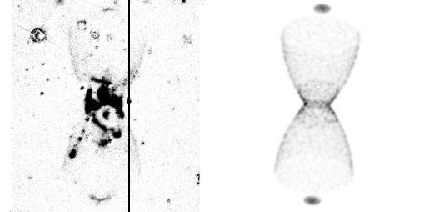}
\caption{\label{bmodel} The H$\alpha$ image (left) and the model (right) of the bipolar nebula of CK\,Vul. Both images are $80 \times 70$ arcsec. North is at the top, East to the left. The vertical line marks the position of the slit.}
\end{figure}

\begin{figure}
\begin{center}
\includegraphics[width=40mm]{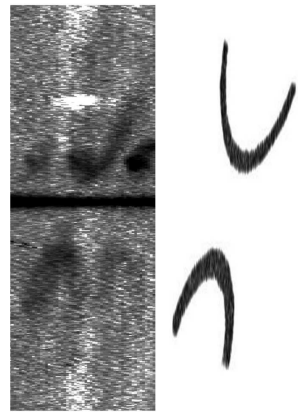}
\end{center}
\caption{\label{pvmodel} The modelled and observed resolved spectrum of the bipolar nebula. The stellar trace of the fast star is in the centre. The spectral coverage is about 60\AA, centered at the wavelength of 6580\AA. The spatial extent is about 80 arcsec.}
\end{figure}

We used the {\sc shape} tool \citep{Steffen10} in order to reproduce the
structure of the bipolar nebula. We used the $\rm H \alpha$ nebular image
(Figure \ref{bmodel}) and the spectrum taken at the position of the fast star
(Figure \ref{pvmodel}). The slit covered a relatively bright part of the bipolar
nebula, unaffected by the bright, central nebula. An inclination angle of 65
degrees was obtained. The modelled and observed image and spectrum of the
bipolar nebula are shown in Figure \ref{bmodel} and \ref{pvmodel}. The obtained
expansion distance of $(0.70 \pm 0.15)$\,kpc is in agreement with the extinction
distance from \citet{Shara85}. The expansion velocity of the bipolar tips, which
are the fastest components of the nebula, is about $\rm 900\,km\,s^{-1}$.

\subsection{The brightness drop of the nebula}

\begin{figure*}
\includegraphics[width=50mm]{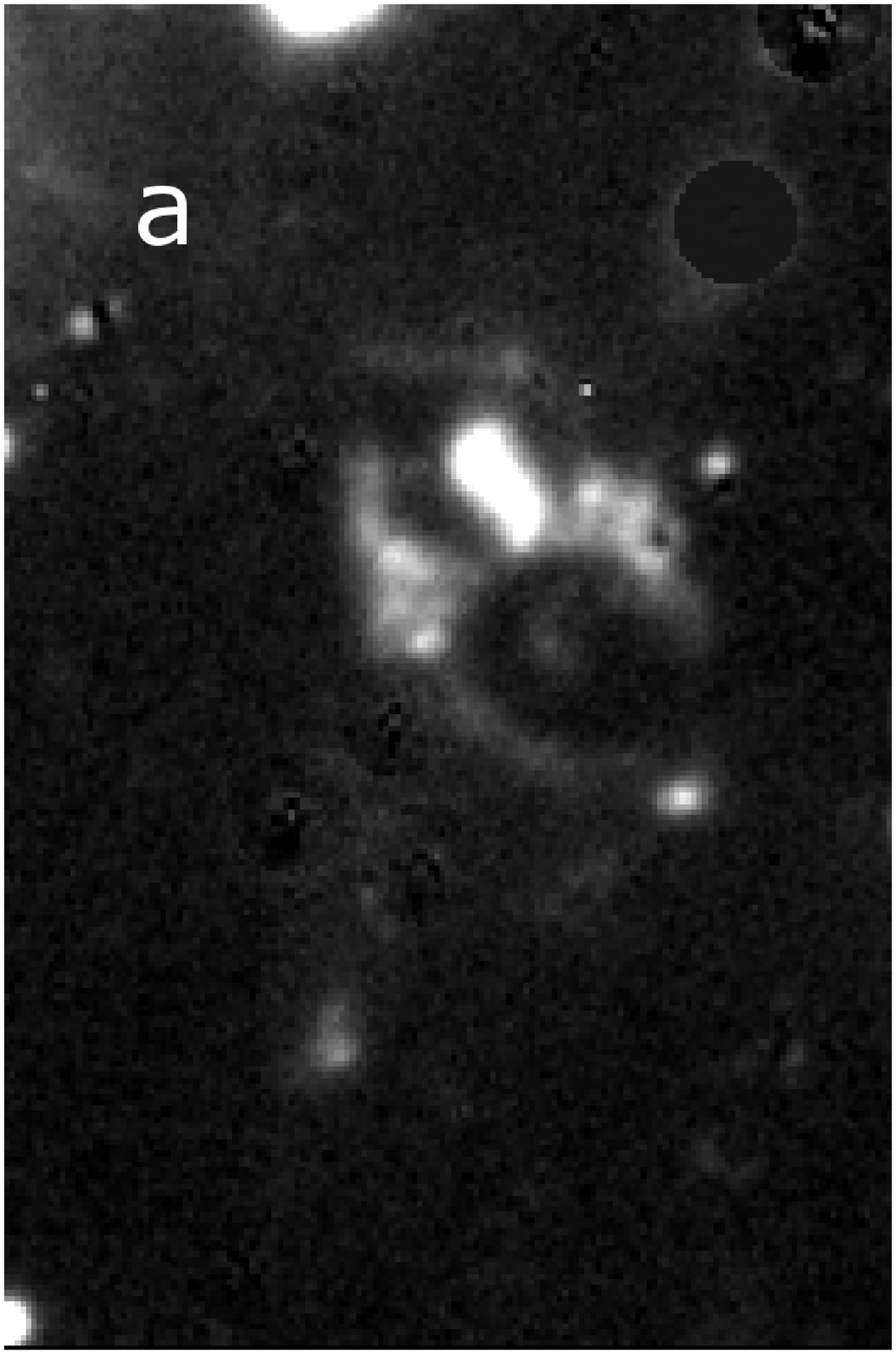}
\includegraphics[width=50mm]{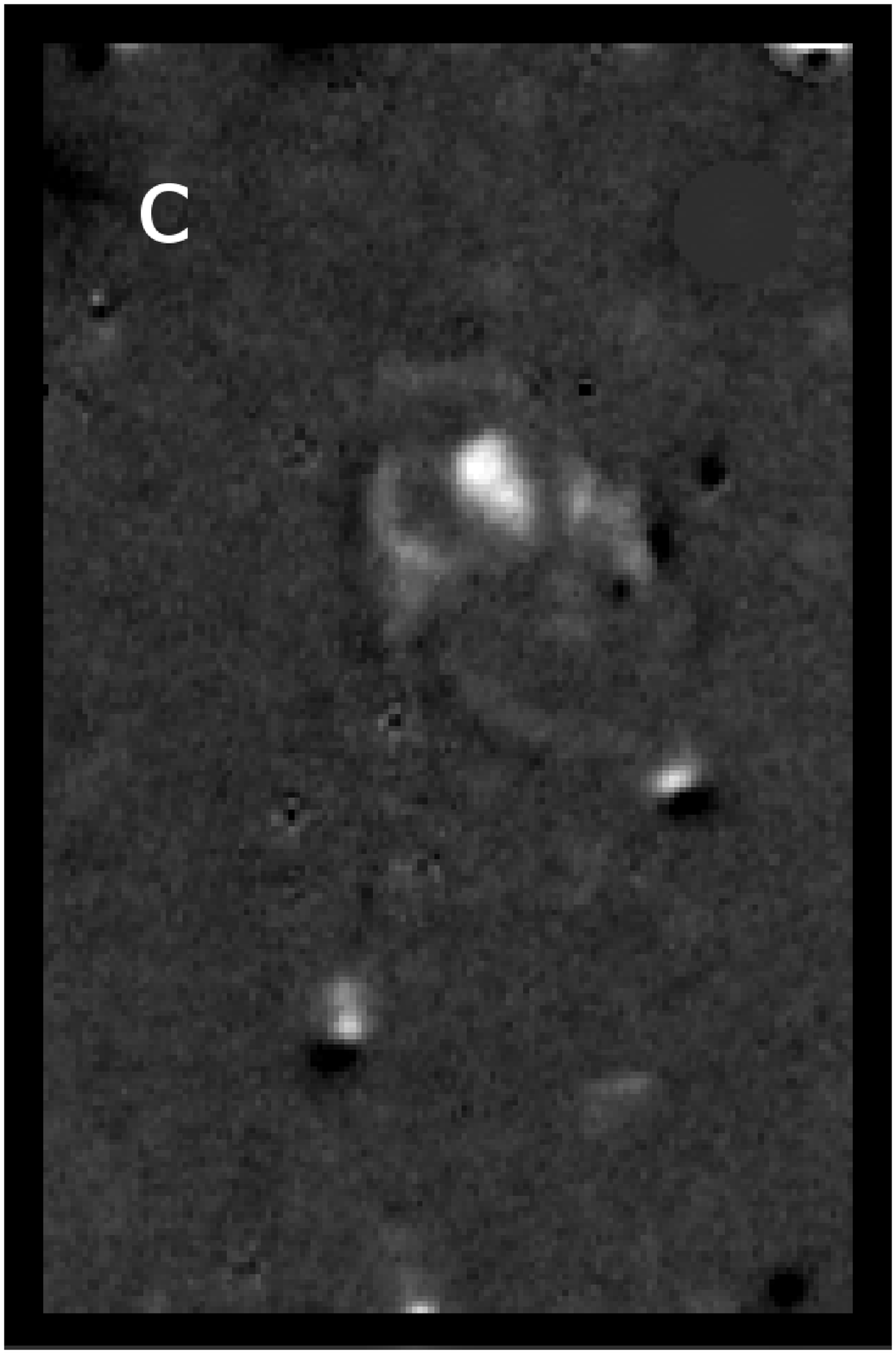}
\includegraphics[width=50mm]{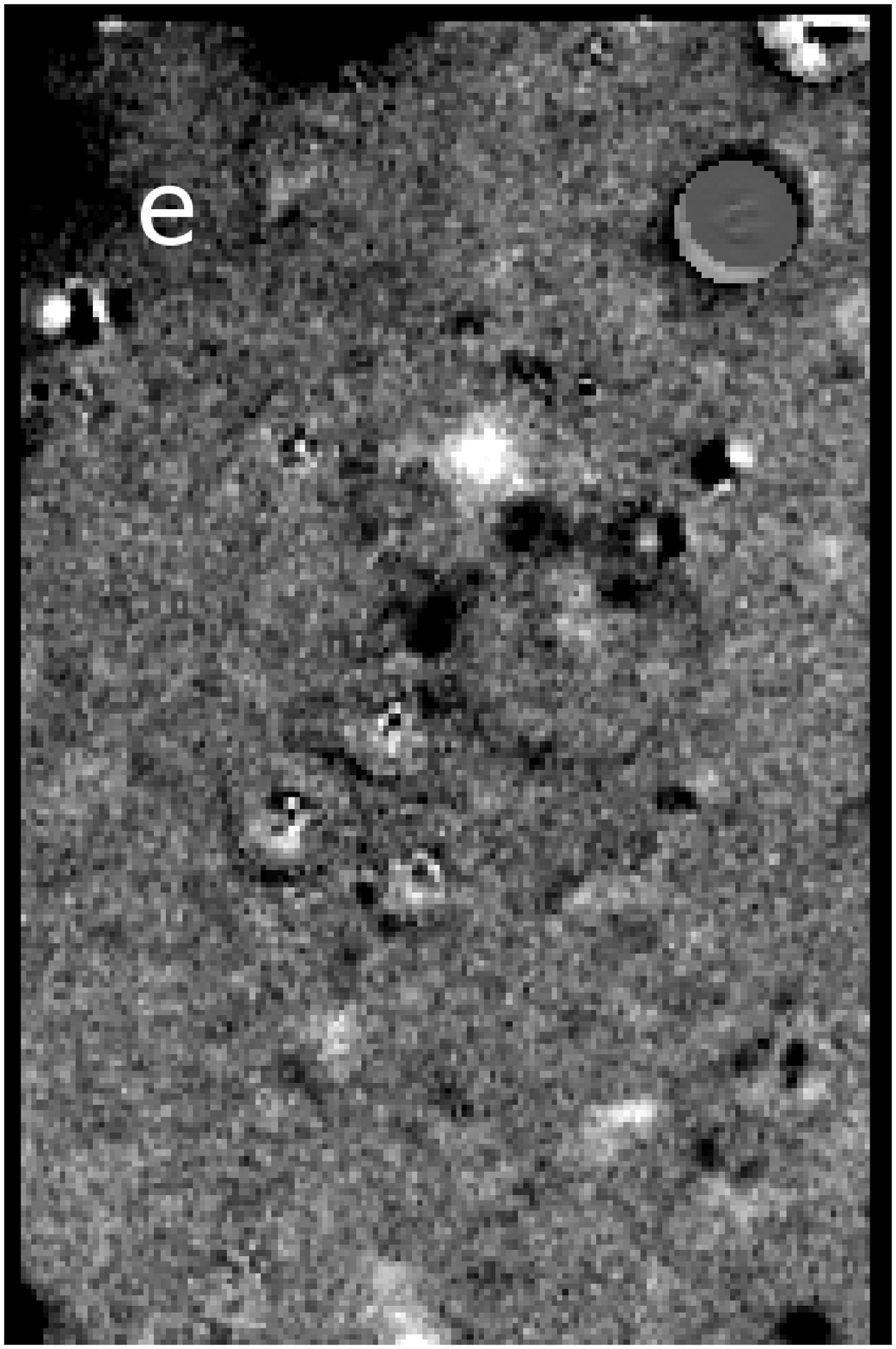}
\includegraphics[width=50mm]{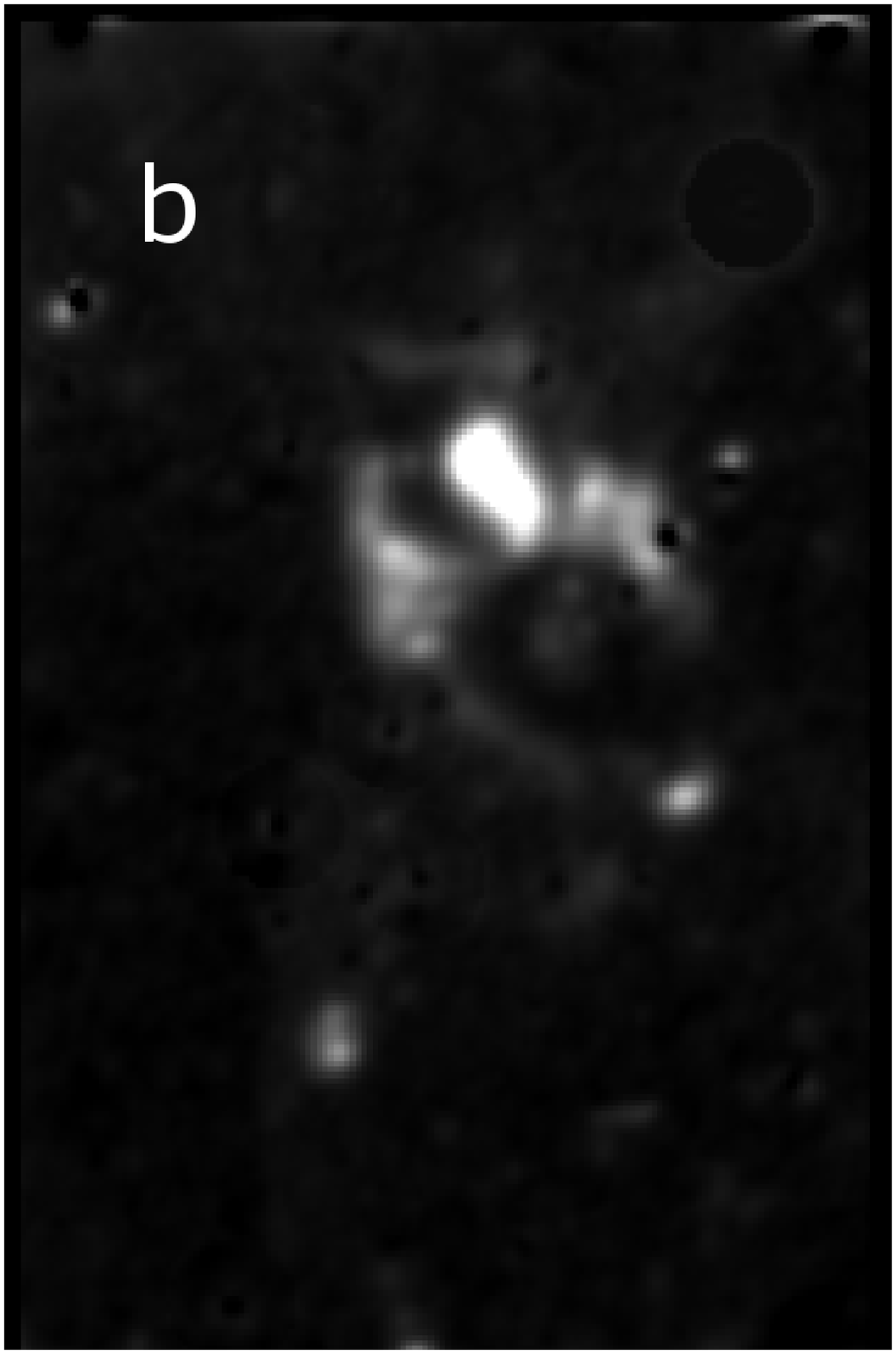}
\includegraphics[width=50mm]{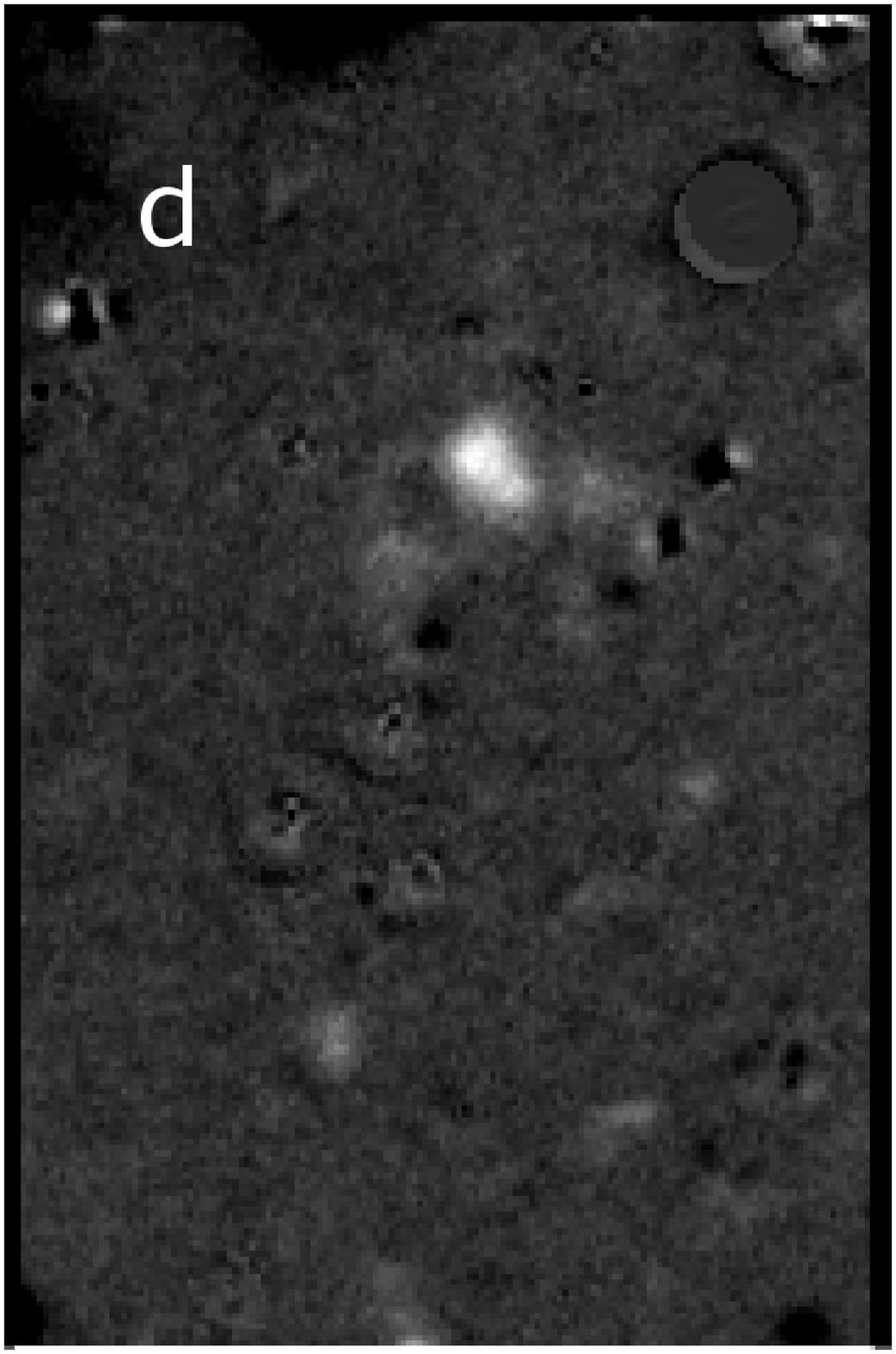}
\includegraphics[width=50mm]{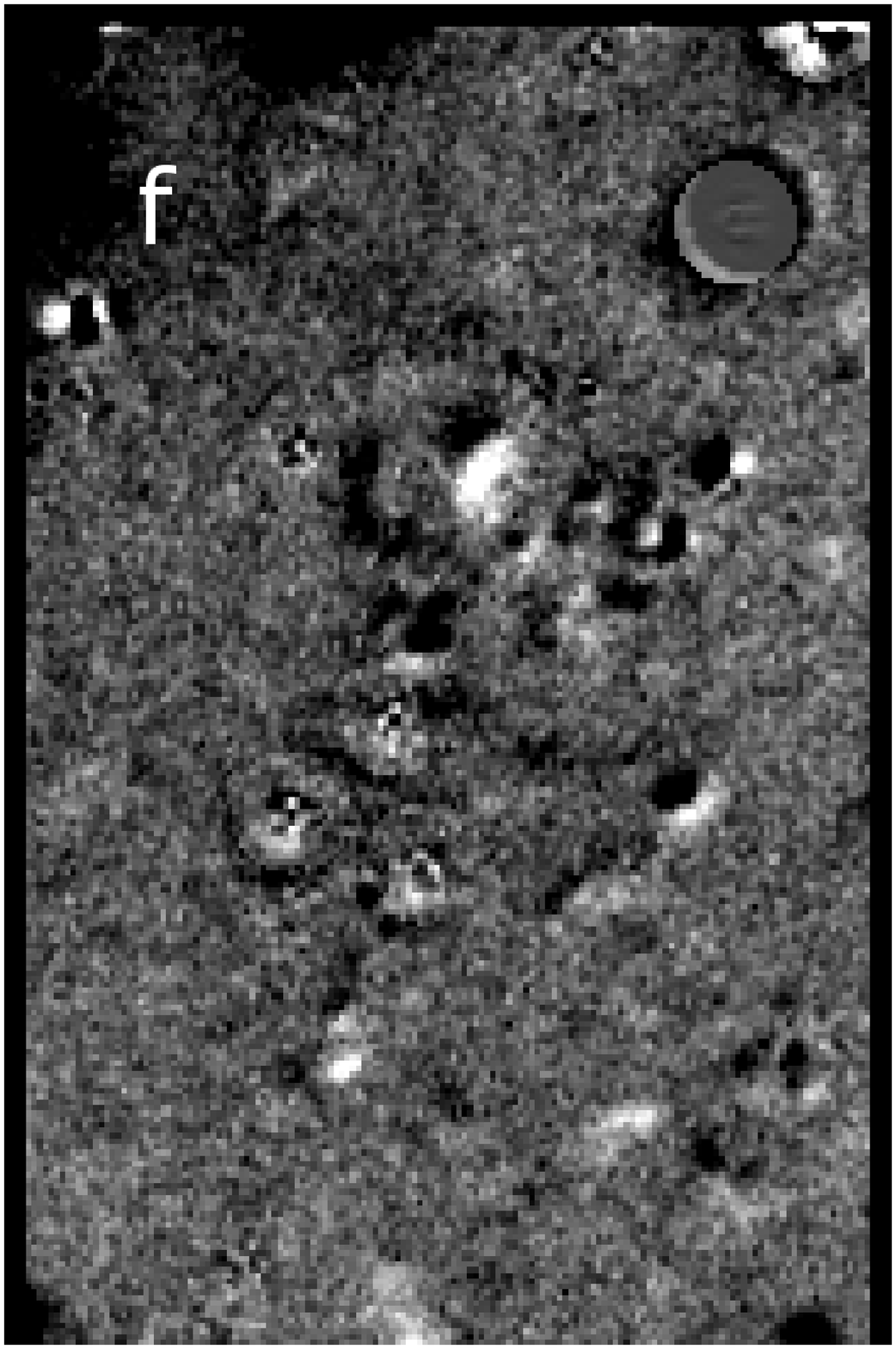}
\caption{\label{dia} H$\alpha$+[{N\sc\,ii}] image of the central ($\rm 40 \times 60 \, arcsec$) region of the nebula made in 2009 and 1991 (a and b, respectively), 1991 minus 2009 and magnified 1991 minus 2009 (c and d), re-scaled H$\alpha$+[{N\sc\,ii}] 1991 image minus 2009 and shifted 1991 minus 2009 (e and f). A star in the upper right corner was used as a PSF model. North is at the top, East to the left.}
\end{figure*}

We used the ISIS subtraction package \citep{Alard98} in order to investigate the
motion and a possible brightess change of the nebula. We have chosen the 1991
and 2009 WHT images for the analysis, since those used a similar H$\alpha$
filter. Both images were spatially aligned and the intensity scale was matched
using the background stars. The PSF of the 1991 image was matched to the PSF of
the 2009 image. Then the background stars were removed. The convolved 1991 and
the 2009 image are shown in Figures~\ref{dia}a and b.

In the next step, the 2009 image was subtracted from the 1991 image. As
expected, a strong residual pattern due to the nebular expansion between 1991
and 2009 was left (Figure \ref{dia}c). In order to compensate for the nebular
expansion, we magnified the 1991 image by a factor of 1.053, corresponding to
the linear expansion of the nebula since 1670, prior to the subtraction. It
reduced the residual pattern caused by the nebular expansion. However, a
residual emission near the center of expansion remained in the new differential
image (Figure \ref{dia}d). It can be attributed to a decrease of the flux of the
$\rm H \alpha$ and [N\,{\sc ii}] lines. The southernmost part of the brightest
blob~1 appears to have remained unchanged, but in the north-eastern part the
flux has dropped significantly between 1991 and 2009. Other parts of the nebula
are also affected by the flux change.

Finally, we checked if this differential pattern could be caused by proper
motion of the nebula with respect to the background stars. We matched the
intensity of the nebula in the 2009 and magnified 1991 images prior to the
subtraction for clarity. The differential image confirms that the flux did not
drop uniformly in the nebula, affecting mostly the north-eastern part of the
blob~1, and blobs 4 and 5 (Figure \ref{dia}e). The south-western part of the
blob~1 is unaffected by the brightness change. The gradient of the brightness
change is parallel to the longest extent of blob~1. The observed differential
pattern could result from the proper motion of the nebula aligned with the
longest extent of blob~1. In order to examine this possibility, we shifted
the 1991 image along the longest extent of blob~1, by 1.0 pix south and 0.5
pix west prior to the subtraction. In the result, the peak of the residual
emission moved to the center of blob~1 (Figure \ref{dia}f). Negative
residuals remained in the south and appeared in the north-eastern part of the
blob~1. Thus, proper motion cannot account for the differential pattern seen in
the residual image.

\subsection{Line ratios}\label{nebula}

We examined the H$\alpha$/[N\,{\sc ii}] 6584\AA\ line ratio in the nebula. There
are no regions where H$\alpha$ remained undetected, indicating lack of H-poor
material in the nebula of CK\,Vul. The H$\alpha$/[N\,{\sc ii}] 6584\AA\ line
ratio ranges from 0.3 to 1, except at the edge of the bipolar nebula, where it
is higher. The H$\alpha$ line exceeds in intensity the [N\,{\sc ii}] 6584\AA\
line only in the bipolar lobes.


Diagnostic lines of [S\,{\sc ii}] and [N\,{\sc ii}] were observed in the nebula.
The density derived from the [S\,{\sc ii}] 6716/6731 \AA\ line ratio indicates
an electron density of $\rm 600\,cm^{-3}$ for blobs~1, 2 and 4. The derived
electron density in the tips of the bipolar nebula is $\rm 200\,cm^{-3}$. The
electron temperature derived from the [N\,{\sc ii}] 5755\AA\ to the [N\,{\sc
ii}] 6548\AA\ + 6584\AA\ line ratio is 12\,000\,K for blob~1 (Table \ref{ratios}).

\begin{table}
 \caption[]{Dereddened and modelled (for $\rm log \, T_{eff} \, [K] = 4.75$ and 5) line intensities in selected parts of the CK\,Vul nebula (relative to $\rm H \beta = 100$).}
 \label{ratios}
 \begin{tabular}{@{}cccccc}
  \hline
  $\lambda_0$ [\AA]&ion		& blob~1	&n. tip	& $\rm log \, T_{eff}=4.75$	& $\rm log \, T_{eff}=5$ \\
  \hline
  4959		&[O\,{\sc iii}]	&94	&	&38	&121	\\
  5007		&[O\,{\sc iii}]	&280	&	&115	&364	\\
  5200		&[N\,{\sc i}]	&69	&	&6	&17	\\
  5755		&[N\,{\sc ii}]	&14	&	&4	&8	\\
  5876		&He\,{\sc i}	&31	&	&15	&15	\\
  6300		&[O\,{\sc i}]	&26	&	&25	&69	\\
  6363		&[O\,{\sc i}]	&10	&	&8	&22	\\
  6548		&[N\,{\sc ii}]	&218	&129	&	&	\\
  6563		&$\rm H \alpha$	&283	&283	&294	&292	\\
  6584		&[N\,{\sc ii}]	&656	&403	&438	&644	\\
  6716		&[S\,{\sc ii}]	&78	&90	&62	&109	\\
  6731		&[S\,{\sc ii}]	&76	&59	&69	&114	\\
  \hline
 \end{tabular}
\end{table}

We determined an interstellar reddening of $E(B-V)=0.7$, using the observed
Balmer decrement and adopting the total to selective extinction ratio $R_V =
3.1$. For the spectra made on June 14th, the derived reddening is much higher
($E(B-V)=1.05$). The synthetic magnitudes of the objects derived on that night
do not agree with the photometry. The sensitivity curve on that night differs
from the sensitivity curves derived on the two other nights. An uncertain
calibration of the spectrum could result from changing weather conditions
between the standard star and target observations. The value of $E(B-V) = 0.7$
obtained on the June 12th is similar to the reddening given by \citet{Shara85}.

The observed drop of the nebular flux may be ascribed to the slowly proceeding
recombination of hydrogen and nitrogen. The timescale of the recombination of
$\rm H^+$ to $\rm H^0$ is 127 years for a density of $\rm 600\,cm^{-3}$
\citep{Nahar97}. The timescale of the recombination of $\rm N^+$ to $\rm N^0$ is
about 100 years at the same density. We observe a drop in the flux by about
20-30\% between 1991 and 2009. If there was no ionizing radiation, the flux
would drop only by 15\% during that period, if charge transfer was not taken
into consideration. Charge transfer between the $\rm N^+$ and $\rm H^0$ becomes
important in a partly neutral medium. It may effectively shorten the timescale
of recombination $\rm N^+$ to $\rm N^0$. 

At very low densities the recombination timescale for hydrogen may be of the
order of a thousand years. Doubly ionized nitrogen may have had not enough time
to recombine and charge transfer is not yet important. This could result in an
enhanced $\rm H$ to [N\,{\sc ii}] line ratio observed in the bipolar lobes.

\begin{figure*}
\includegraphics[width=160mm]{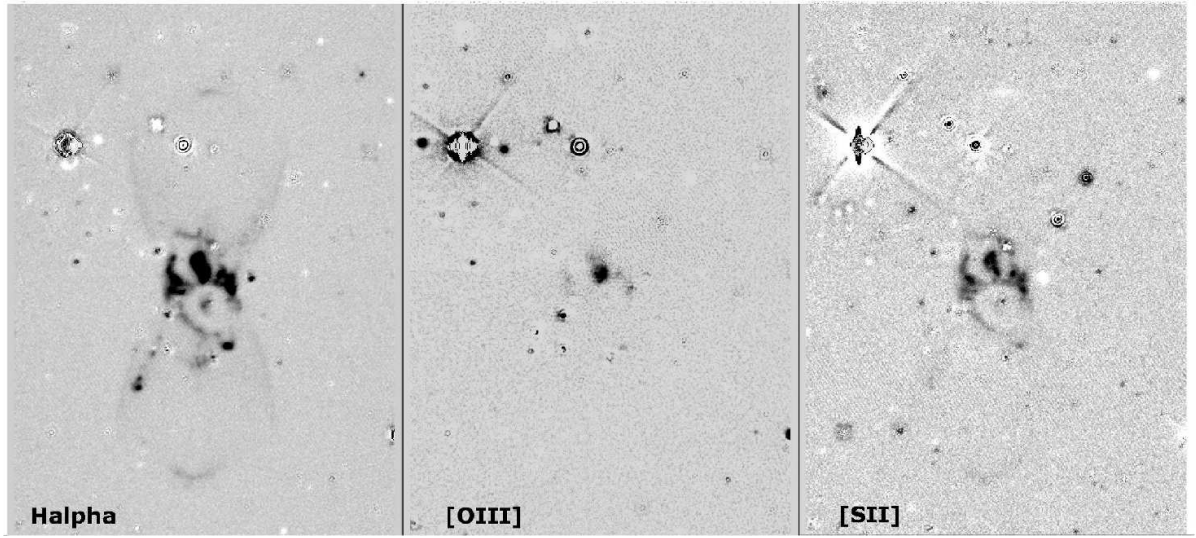}
\caption{\label{emission} Gemini continuum subtracted $\rm H\alpha$, [O{\sc\,iii}] and [S{\sc\,ii}] images. Each image is about $70 \times 100$ arcsec. North is at the top, East to the left.}
\end{figure*}

The central source appears to be still active and to maintain the ionization of
the innermost parts of the nebula, in particular the southern part of
nebulosity~1, not affected by the brightness drop. The [O\,{\sc iii}] 5007\AA\
emission is concentrated near the presumed position of the central source
(Figure \ref{emission}). The recombination timescale for $\rm O^{++} \rightarrow
O^+$ at $n_e = 600$ is only 21 years, ignoring charge transfer \citep{Nahar97}.
An $\rm O^{++}$ sphere of the radius of $\rm 3 \times 10^{16}\,cm$ (about 3
arcsec at the distance to CK\,Vul) would require continuous ionization by the
central source, e.g. characterized with $\rm L \approx 3 L_{\odot}$ and $\rm
T_{\star} = 60000\,K$. Radio observations confirm, that the central source is
still active, although its luminosity may be gradually decreasing
\citep{Hajduk07}.

We run {\sc cloudy} models in order to reproduce the line intensity ratios in
the innermost part of the nebula (which does not show flux change in time).
Table \ref{ratios} presents line ratios obtained for the model.








We computed the ionized mass using the image and spectra. We approximated the
1st blob with a filled cylinder of 6 arcsec height, 1.5 arcsec radius and the
mean proton density of $\rm 600 \, cm^{-3}$, and the 2nd and 3rd blobs with
filled cones with densities of 300 and $\rm 600 \, cm^{-3}$ and radius and
height of 5 and 6.5 arcsec, respectively. The total ionized mass of the central
nebula is about $6 \times 10 ^{-5} M_{\odot}$. The mass of the limb-brightened
bipolar nebula may be one order of magnitude higher. This is two orders of
magnitude less than the mass of the nebula estimated in \citet{Hajduk07}, who
assumed uniform nebular density throughout the bipolar nebula.

\section{The fast star in the vicinity of CK\,Vul}

\subsection{Proper motion of the star and the nebula}

\begin{figure}
\includegraphics[width=84mm]{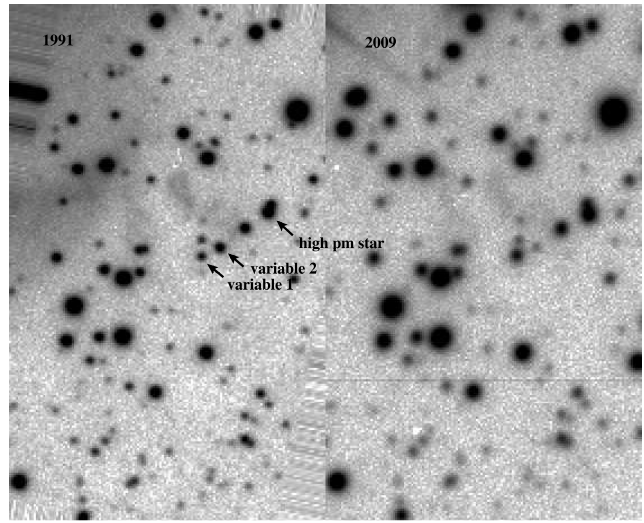}
\caption{\label{var} Comparison of the 1991 and 2009 $R$ band images. The high proper motion and two variable stars are marked with arrows. North is at the top, East to the left. The image scale is $60 \times 40$ arcsec.}
\end{figure}

The $R$ band images taken in 1991 and 2009 are presented in Figure \ref{var}.
One of the stars in the vicinity of CK\,Vul is characterized by high proper
motion. The direction of the motion of the star appears to point away from the
center of the nebula. We attempted to verify a possible link of the high proper
motion star with CK\,Vul.

\begin{figure}
\includegraphics[width=84mm]{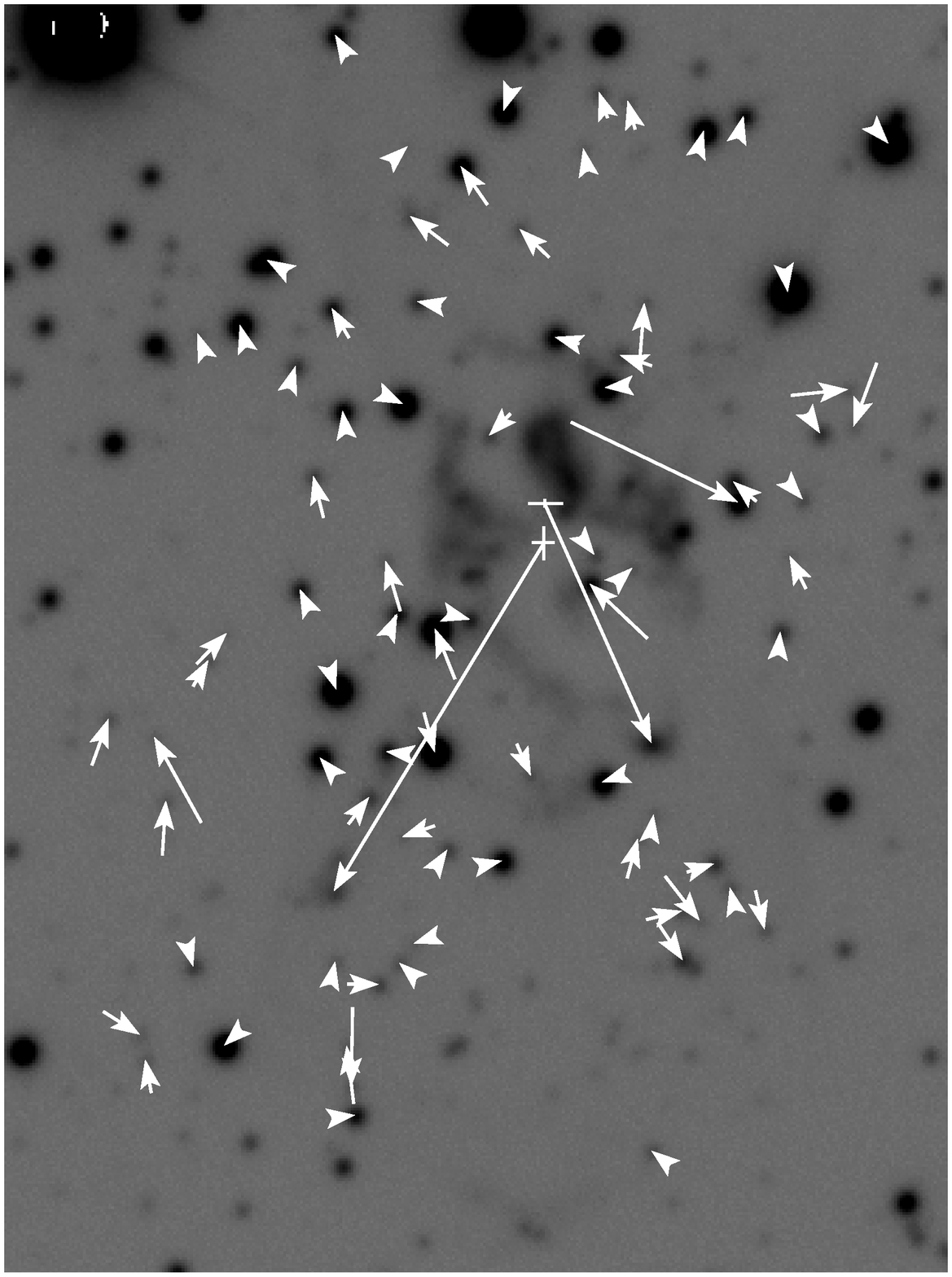}
\caption{\label{pm} The Gemini H$\alpha$ image of the field of CK\,Vul. The
proper motions represented by arrows derived from the 1991 and 2010 images are
exaggerated by a factor of 17.9, corresponding to the movement since 1670.
A logarithmic flux scale is used. North is at the top, East to the left. The image scale is $66 \times 47$ arcsec.}
\end{figure}

The Gemini image of CK\,Vul taken through the $\rm H \alpha $ filter is
presented in Figure \ref{pm}. We fitted stellar positions in the 2010 and 1991
images using the {\sc iraf daophot} package. Arrows mark extrapolated proper
motions of the stars between 1670 and 2010. The mean proper motion of all the
field stars present in both images (excluding the three stars for which
association with CK\,Vul is investigated), serves as the reference frame. 

In addition, we measured the proper motions of blob~4 and 5. They are marked in
Figure \ref{pm} along with the error bars. We measured the centroid positions,
since both of the blobs are extended and cannot be fitted with a stellar PSF.
The proper motions measured for the nebular blobs are less accurate than for
stars. The error bars are derived from the rms of the centroid positions
measured using different radii.

\begin{figure*}
\includegraphics[width=150mm]{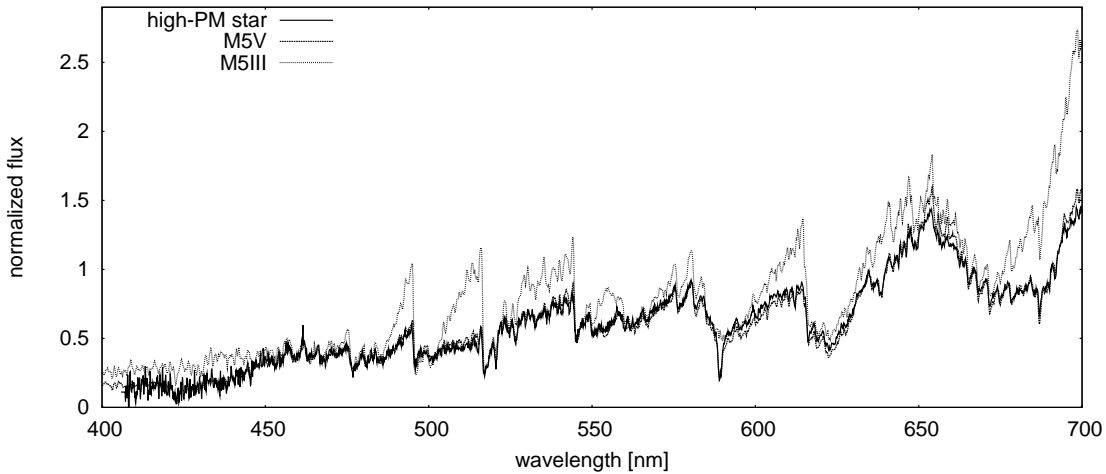}
\caption{\label{spect1b} Spectrum of the high proper-motion
star near CK\,Vul. Spectra of M5V and M5III type stars are plotted for comparison. The spectra were matched in intensity.}
\end{figure*}

The high proper motion star is characterized by $\rm \mu = 1000\,mas \, yr^{-1}$
at a position angle of 244.4 degree (or $\rm \mu_{\alpha} = -900\,mas \,
yr^{-1}$ and $\rm \mu_{\delta} = -400\,mas \, yr^{-1}$). The extrapolated
position of the fast star is a few arcseconds north of the position of the
nebula determined for 1670, when the outburst occured. The accuracy of the
position determination of the star on both images may be questioned due to
blending with a nearby field star. However, the PSF fitting algorithm was able
to resolve both stars. The residuals in the PSF subtracted image assure us that
the close blend did not influence the position measurements for the fast star. 

The different positions of the star and the nebula extrapolated to 1670 argue
against a connection of the high proper motion star with CK\,Vul. However, we do
not find the positional measurements decisive. The proper motion of the nebula
may be affected by different seeing, sampling or/and brightness changes of the
nebula between 1991 and 2010. The possible link of CK\,Vul and the fast star is
investigated again on the basis of photometric and spectroscopic observations of
the star and the nebula in the following subsection.

\subsection{Photometry and spectroscopy of the fast star}

The spectrum of the fast star is presented in Figure~\ref{spect1b}. It is
compared to an M5 type dwarf and giant spectra, taken from \cite{Jacoby84}. The
fast star resembles a dwarf rather than a giant. It does not show either TiO
absorptions as deep as a giant, or the $\rm H \alpha$ absorption. The sodium D
doublet, strong in the observed star, is likely of stellar origin. The CaH1
spectral index of 0.84 \citep{gizis97} indicates that the star is not a
subdwarf.


The observed brightness of the high proper motion star is $R \approx 18.5$
magnitude. Stellar type and extinction were fitted to minimalize the discrepancy
between the synthetic and observed $g-r, r-i$ and $i-z$ colors \citep{covey07}.
The best fit was obtained for an M4 type with no additional extinction. The
distance of an M4 star with the apparent $r = 18.85$ is ($240 \pm 40$)\,pc,
derived from the color-luminosity relation \citep{Juric08}. If we used the M5
type, derived from the spectroscopy, the distance would shrink to ($140 \pm
40$)\,pc.

The spectrum of the high proper-motion star in Figure \ref{spect1b} was not
corrected for interstellar extinction. The comparison with the spectrum of the
unreddened M dwarf taken from Jacoby's library demonstrates that the fast star
is not reddened, contrary to CK\,Vul. Both the distance and the interstellar
reddening indicate, that the fast star is a foreground star, located much closer
than CK\,Vul. The tangential velocity of the star is 330\,km/s for the distance
of 140\,pc. The heliocentric radial velocity of the star is $\rm
-50\,km\,s^{-1}$.

The kinematical properties of the fast star are not consistent with the galactic
disk. Such deviations are frequently observed in the solar neighbourhood for
non-metal-poor dwarfs. \citet{lepine03} discuss possible mechanisms explaining
those deviations.

\section{Variable stars}

\subsection{Brightness evolution of the variable stars}

Two stars located within $3-4$ arcsec of the expansion centre in the plane of
the sky show pronounced variability (Figure \ref{var}). The apparent separation of the
two variable stars is about 2 arcsec.

\begin{figure}
\includegraphics[width=84mm]{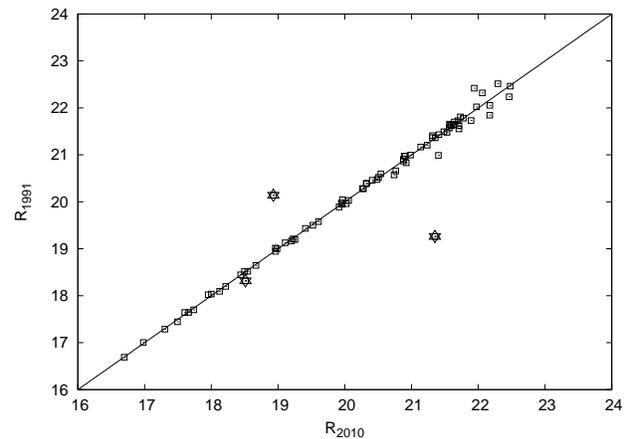}
\caption{\label{phot} Comparison between 2010 June 23 and 1991 August 10 $R$ band magnitudes of stars in the field of CK\,Vul. Three stars in the vicinity of the expansion centre, characterized with high proper motion or magnitude change are marked.}
\end{figure}

The comparison of the brightness of the field stars between 2010 and 1991 is
presented in Figure \ref{phot}. The $g'r'i'z'$ photometry was transformed to R
magnitudes for the 2010 Gemini image \citep{Smith02}. Two of the field stars showing significant brightness changes and the high proper motion star are marked with a different symbol.

\begin{figure}
\includegraphics[width=84mm]{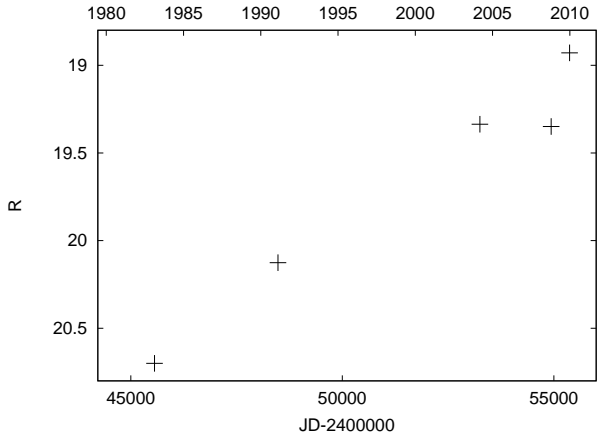}
\caption{\label{phot2} Evolution of the $R$ band brightness of the 1st variable
star in the field of CK\,Vul. The sloan $r'$ observations have been
transformed to Johnson $R$.}
\end{figure}

\begin{figure}
\includegraphics[width=84mm]{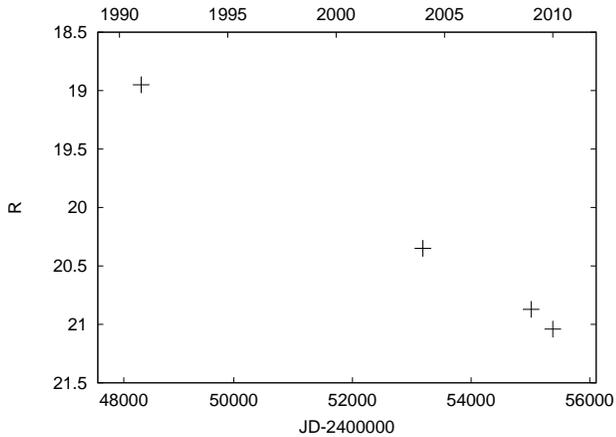}
\caption{\label{phot3} Evolution of the $R$ band brightness of the 2nd variable 
star in the field of CK\,Vul. The sloan $r'$ observation have been
transformed to Johnson $R$.}
\end{figure}

The lightcurves of the two variable stars between 1991 and 2010 are shown in Figures~\ref{phot2} and \ref{phot3}. One of the variable stars (referred to as
the 2nd variable) became significantly fainter during the observing period,
while the other one (1st variable) has brightened. \cite{Shara85} considered the
latter star as a central star candidate and estimated its brightness as $R =
20.7 \pm 0.5$ in August~1983. 

The observed decline of the 2nd variable continues for at least 20 years, and
the increase of the brightness of the 1st variable for 30 years. We were able to
measure the colour change for the 1st variable using the IPHAS 2005 and GMOS
2010 $r'i'$ observations. The $r'-i'$ index of variable~1 was equal to 1.73 in
2005 and to 1.61 in 2010 (in the Vega system), thus it changed by $\Delta
(r'-i') = - 0.12 \pm 0.08$ during\,5 years. The $r'$ magnitude decreased by
$0.20 \pm 0.05$ magnitude. For dust characterized by $R_V=3.1$, the brightness
change between 1991 and 2005 of $\Delta r' = 0.2$ would imply a change of the
index $\Delta (r'-i')$ of $ - 0.05$ \citep{cardelli89}. The observed colour
change of $\Delta (r'-i') = - 0.12$ suggest low value of the $R_V$, although
this requires further confirmation.

Observed photometric indicies of the 1st variable are best fitted with the
synthetic colors of a K4 main sequence star and $A_V = 4.4$ or
an M0 star with $A_V = 3.3$. Stellar type later than M0 can be ruled out, though
earlier type than K4 is possible. The 2nd variable is best fitted with a K4 star
and  $A_V = 3.7$. The distances to variable~1 obtained using the photometric
parallax method \citep{Juric08} are ($540 \pm 100$)\,pc for a K4 star and
$A_{r'} = 3.7$ and ($440 \pm 80$)\,pc for an M0 star and extinction of $A_{r'} =
2.8$. The calculated distance to variable~2 is ($2000 \pm 400$)\,pc, assuming
extinction of $A_{r'} = 3.1$.

The obtained extinction and spectral types must be treated with caution. We
assumed the variables to be main sequence stars and the Galactic extinction law
with $R_V = 3.1$.

\subsection{Spectra of the variable stars}

\begin{figure}
\includegraphics[width=84mm]{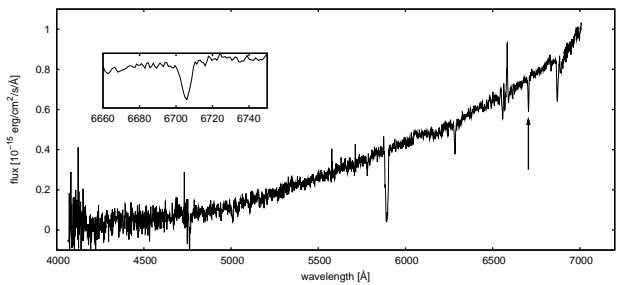}
\caption{\label{spect2} Spectrum of variable 1. An arrow marks the 6705~\AA\ line. An inset presents vicinity of the 6705~\AA\ line.}
\end{figure}

\begin{figure}
\includegraphics[width=84mm]{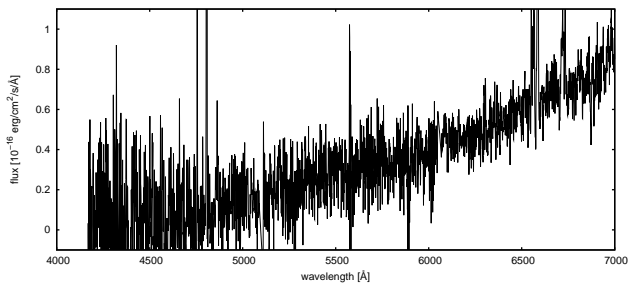}
\caption{\label{spect3} Spectrum of variable 2.}
\end{figure}

The spectra of both variables are presented in Figures~\ref{spect2} and
\ref{spect3}. The spectrum of variable~1 has much better quality than that of
variable~2, due to difference in brightness of both stars. No stellar lines are
identified in the spectrum of variable~1, apart from the $\rm H \alpha$
absorption (which is difficult to confirm because of the superposition of the
star and blob~6) and the stellar contribution to the Na\,{\sc I} 5890 and
5896\AA\ absorption. Variable~1 was reported to show $\rm H \alpha$ absorption
by \citet{Naylor92}, although their spectrum had worse quality from that ours.

Neither spectrum shows molecular bands. An M0 or later type star can be ruled
out on the basis of the shape of the stellar continuum. In order to match the
spectrum of variable~1 to a spectrum of a K4V star, we need to apply the
reddening of $A_V \approx 5.7$ - much higher than from photometric observations
($A_V = 4.4$). This may arise from uncertain calibration of the spectroscopic
observations (see section \ref{nebula}).

We detected a saturated Na\,{\sc I} 5890 and 5896\AA\ doublet of interstellar
origin (with a possible contribution of the stellar absorption) in both spectra
(Figure~\ref{spect2}). In addition, several diffuse interstellar bands (DIB) are
superimposed on the continuum of the 2nd variable. The strongest one, with an
equivalent width (EW) of 2\,\AA, is the 6284\,\AA\ feature. Other fainter DIBs
are present at 5781, 5798 and 6613\,\AA. Those features are not detectable in
the spectrum of variable~2 due to the poor S/N ratio.

A strong ($ \rm EW \approx 1 $\,\AA) absorption is seen at 6705~\AA\ in the
spectrum of variable~1. It is also present (and possibly stronger) in the
spectrum of variable~2. We identify this feature as a resonance Li\,{\sc i}
line. 

A stellar origin of the 6705\,\AA\ line is doubtful. Lithium is depleted in low
mass stars before they reach the main sequence phase, so the two stars would
have to be very young to show strong 6705~\AA\ absorption. We do not find any
evidence that the two stars reside in a star forming region or an open cluster.
The two stars appear to lie at different distances along the line of sight. The
EW of the Li\,{\sc i} line of 1\AA\ would require one of the highest lithium
overabundances observed in stars \citep{Soderblom93}.

An alternative identification for the absorption line could be a Ce\,{\sc ii}
transition \citep{Reyniers02}, observed in atmospheres of post AGB-stars. None
of the variables appears to be an evolved star. No other stellar lines or
molecular bands are detected in the spectra, while the 6705\,\AA\ line is very
strong.

An interstellar origin of this line is also unlikely. Only very weak Li\,{\sc i}
absorption is observed in the sightline towards a handful of stars, e.g. $\rm
\zeta$\,Oph \citep{Howarth02}. 


The variable extinction and the Li\,{\sc i} line most likely originate in the
same medium in the sightline towards the two stars.  The stars could be partly
obscured by the matter ejected in the 1670 explosion, similar to the blue
component in V838\,Mon \citep{Tylenda11}. That produced a prominent absorption
of Li\,{\sc i} and other elements in the spectrum of V838\,Mon. It is doubtful,
however, that the stars are physically associated with CK\,Vul. The distance to
variable~2 appears to be much larger than that of CK\,Vul, thus it is likely a
background star. Variable 1 is close to the distance to CK\,Vul. Within
uncertainties, it can be a located behind CK\,Vul.

\subsection{Dusty cloud ejected by CK\,Vul}

Although the two variable stars are identyfied as background objects, their
spectra and photometry can be used to investigate the properties of the cloud of
gas and dust ejected by CK\,Vul. The cloud must extend at least $\sim 200$\,AU
in the direction to the expansion centre (assuming that the cloud was ejected in
1670), since both stars remain variable for at least 20 to 30 years. 

In order to determine the extinction caused by the cloud, we have to disentangle
it from the interstellar extinction. The interstellar medium forms two
extinction layers towards CK\,Vul: the first one at a distance of 500\,pc and
the other one at 2\,kpc. An object located between the two layers would be
reddened by $E(B-V) \simeq 0.8$ and behind the second layer by $E(B-V) \simeq
1.25$ \citep{Weight03}. The distance to CK\,Vul of 0.70\,kpc indicates that the
object lies between the two extinction layers observed in its direction. Most,
if not all, of the reddening suffered by the nebula of CK\,Vul is of
interstellar origin. 

The lower limit for the extinction for variable~2 is $A_R = 2$, assuming that
the star was not obscured by the cloud in 1990, when the first observation was
made. This would require higher extinction than observed in the V band and may
be due to uncertainty in the stellar type determination or/and unusual dust
properties.

If the same cloud is responsible for the variability of both stars, then its
extent is at least 1650\,AU (corresponding to separation of the stars of 2.2
arcsec at the distance of CK\,Vul). The observed EW of the line of 1\,\AA\
corresponds to a neutral lithium column density of $\rm 5 \times
10^{12}\,cm^{-2}$.

Assuming the extent of the cloud is of the order of the projected distance of
the two stars, the total lithium mass is of the order of $\rm 2 \times 10^{-11}
M_{\odot}$. The derived velocity of the cloud is about $\rm 100 \, km \,
s^{-1}$, if it expands in a plane of the sky.

The dusty clump may be identified with the cold component seen in the SED of
CK\,Vul \citep{Evans02}, while the hotter component may be associated with the
hotter dust obscuring the central star. The total mass of the dusty clump may be
equal to $\rm 2 \times 10 ^{-2} M_{\odot}$, previously attributed to the bipolar
nebula \citep{Hajduk07}.

\section{Discussion}

Discussion of the nature of CK\,Vul, which continues since its re-discovery by
\citet{Shara82}, is impeded by observational difficulties and apparent lack of
objects of similar characteristics. The central source was detected only in the
radio domain and appears to remain (partially?) obscured. Its temperature and
luminosity are high enough to balance the recombination of $\rm O^{+}$ in its
vicinity. The geometry of the obscuring body is unknown \citep{Hajduk07}. We can
only very roughly constraint the central source luminosity and temperature. The
luminosity of a few solar luminosities and a temperature of 60\,000\,K would
result in a very small diameter of the emitting body (less than 1000\,km).
However, in the case of much higher luminosity the source would be able to
ionize all the nebula or, if obscured, would be much more luminous source in the
infrared.

An accreting white dwarf is heated to a temperature of 10-20kK. A thermonuclear
runaway will heat the white dwarf but this is short-lasting as only the surface
is heated - after the runaway, the star goes back to its previous temperatures.
Using equations of \citet{townsley09}, a temperature of 60kK corresponds to an
accretion rate of $\rm 10^{-8} \, M_\odot \, yr^{-1}$. At this \.M, the
luminosity of the white dwarf is about $\rm 0.5 \, L_\odot$. (It depends on the
mass of the white dwarf). The thermal time scale (the rate over which the star
cools down if the accretion stops) is 200 yr (for an $\rm 0.9 \, M_\odot$ white
dwarf). So if the thermonuclear runaway in 1670 AD ended a phase of such high
accretion, the current parameters are not implausible. Or the accretion may
still continue at this rate. MV Lyr is a known CV with parameters similar to
this.

The remnants of the object may be enhanced in Li. The velocities of the ejected
envelope range from tens to almost $\rm 1000\,km\,s^{-1}$, comprising both
symmetric and nonsymmetric components. A massive cloud of dust and gas was
ejected.


The model of a light nova can qualitatively reproduce the lightcurve of CK\,Vul
\citep{Hajduk07,Miller11}. In the nova scenario, the lithium could have been
produced during the outburst of CK\,Vul \citep{Dantona10}. However, the nova
scenario cannot account for other properties of CK\,Vul. The infrared SED of
CK\,Vul shows two maxima corresponding to dust shells of 550\,K and 25\,K
\citep{Evans02}, which are hard to explain in the framework of the nova
explosion. The nova scenario is also at odds with the large mass and the
observed range of velocities of the expelled material.



The total energy radiated by CK\,Vul during its outburst of about $\rm
10^{47}\,erg$, places this object close to NGC\,3432\,OT2008-9 on the
energy-time diagram for observed transients \citep{Soker11}. \citet{Soker11}
propose that outbursts of intermediate luminosity optical transients may be
powered by mass accretion onto a main-sequence companion. It can explain the
expansion velocity of the fastest parts of the CK\,Vul nebula (of the order of
the escape velocity from a MS star), though most of the matter was expelled with
lower velocities. The small radius of the ionizing source is at odds with this
scenario. The white dwarf would not cool as fast as in CK\,Vul, unless the
outburst occured on the surface of the star.

\citet{Kato03} proposed a merger scenario. Such a scenario can reproduce the
variety of light curves and the amplitude of the outburst \citep{Tylenda06}. One
of the merger components should have been lithium rich to explain the lithium
enhancement of the remnants (e.g. a planet, young star or an RGB/AGB star). The
temperature of the source in the expansion centre suggests that one of the
objects was or had become a white dwarf after the outburst. Again, the outburst
would have to occur on the surface of the star and not affect the deeper layers.



The born-again scenario \citep{Evans02} would be in line with the observed
lithium overabundance. However, the lack of hydrogen-deficient material in the
nebula does not agree with the evolutionary models for VLTP. Neither the diffuse
induced nova nor VLTP events can account for the very low luminosity of CK\,Vul
only 340 yr after the outburst \citep{Miller11}.



\section*{Acknowledgments}

The research reported in this paper has partly been supported by a grant No.
NN203 403939 financed by the Polish Ministery of Sciences and Higher Education.
PvH acknowledges support from the Belgian Science Policy office through the
PRODEX program. Based on observations obtained at the Gemini Observatory, which
is operated by the  Association of Universities for Research in Astronomy, Inc.,
under a cooperative agreement  with the NSF on behalf of the Gemini partnership:
the National Science Foundation (United  States), the Science and Technology
Facilities Council (United Kingdom), the  National Research Council (Canada),
CONICYT (Chile), the Australian Research Council (Australia),  Minist\'{e}rio da
Ci\^{e}ncia, Tecnologia e Inova\c{c}\~{a}o (Brazil)  and Ministerio de Ciencia,
Tecnolog\'{i}a e Innovaci\'{o}n Productiva (Argentina). The William Herschel
Telescope and its service programme are operated on the island of La Palma by
the Isaac Newton Group in the Spanish Observatorio del Roque de los Muchachos of
the Instituto de Astrofísica de Canarias. This paper makes use of data obtained
from the Isaac Newton Group Archive which is maintained as part of the CASU
Astronomical Data Centre at the Institute of Astronomy, Cambridge.


\begin{thebibliography}{99}

\bibitem[\protect\citeauthoryear{Alard \& Lupton}{1998}]{Alard98} Alard C., 
Lupton R. H., 1998, ApJ, 503, 325


\bibitem[\protect\citeauthoryear{Cardelli et al.}{1989}]{cardelli89} Cardelli J. A., Clayton G. C., Mathis J. S., 1989, ApJ, 345, 245

\bibitem[\protect\citeauthoryear{Covey et al.}{2007}]{covey07} Covey K. R., et al., 2007, AJ, 134, 2398

\bibitem[\protect\citeauthoryear{D'Antona \& Ventura}{2010}]{Dantona10} D'Antona F., Ventura P., 2010, Proc. IAU Symp., 268, ed. C. Charbonnel, et al., 395

\bibitem[\protect\citeauthoryear{Drew et al.}{2005}]{drew05} Drew J., et al., 2005, MNRAS, 362, 753

\bibitem[\protect\citeauthoryear{Evans et al.}{2002}]{Evans02} Evans A., van Loon J. Th., Zijlstra A. A., Pollacco D., Smalley B., Tyne V. H., Eyres S. P. S., 2002, MNRAS, 332, 35


\bibitem[\protect\citeauthoryear{Fukugita et al.}{1995}]{fukugita95} Fukugita M., Ichikawa T., Gunn J. E., Doi M., Shimasaku K., 1995, AJ, 1996, 111, 1748


\bibitem[\protect\citeauthoryear{Gizis}{1997}]{gizis97} Gizis J. E., 1997, AJ, 113, 806

\bibitem[\protect\citeauthoryear{Hajduk et al.}{2007}]{Hajduk07} Hajduk M., 
et al., 2007, MNRAS, 378, 1298

\bibitem[\protect\citeauthoryear{Harrison}{1996}]{Harrison96} Harrison T. E.,
1996, PASP, 108, 1112

\bibitem[\protect\citeauthoryear{Howarth et al.}{2002}]{Howarth02} Howarth I. D., Price R. J., Crawford, I. A., Hawkins, I., 2002, MNRAS, 335, 267

\bibitem[\protect\citeauthoryear{Jacoby et al.}{1984}]{Jacoby84} Jacoby G. H., Hunter D. A., Christian C. A., ApJSS, 56, 257

\bibitem[\protect\citeauthoryear{Juri\'{c} et al.}{2008}]{Juric08} Juri\'{c} M., et al., 2008, ApJ, 673, 864

\bibitem[\protect\citeauthoryear{Kato}{2003}]{Kato03} Kato T., 2003, 
A\&A, 399, 695

\bibitem[\protect\citeauthoryear{L\'{e}pine et al.}{2003}]{lepine03} L\'{e}pine S., Rich R. M., Shara M. M., 2003, AJ, 125, 1598


\bibitem[\protect\citeauthoryear{Miller Bertolami et al.}{2011}]{Miller11}
Miller Bertolami M. M., Althaus L. G., Olano C., Jim\'{e}nez N., 2011, MNRAS, 415, 1396

\bibitem[\protect\citeauthoryear{Nahar \& Pradhan}{1997}]{Nahar97} 
Nahar S. N., Pradhan A. K., 1997, ApJS, 111, 339

\bibitem[\protect\citeauthoryear{Naylor et al.}{1992}]{Naylor92} Naylor T., 
Charles P. A., Mukai K., Evans A., 1992, MNRAS, 258, 449

\bibitem[\protect\citeauthoryear{Reyniers et al.}{2002}]{Reyniers02} Reyniers M., Van Winckel H., Bi\'{e}mont E., Quinet P., 2002, A\&A, 395, 35

\bibitem[\protect\citeauthoryear{Shara \& Moffat}{1982}]{Shara82} 
Shara M. M.,  Moffat A. A., 1982, ApJ, 258, 41

\bibitem[\protect\citeauthoryear{Shara et al.}{1985}]{Shara85} Shara M. M., 
Moffat A. A., Webbink R. F., 1985, ApJ, 294, 271

\bibitem[\protect\citeauthoryear{Smith et al.}{2002}]{Smith02} Smith J. A., et 
al., 2002, AJ, 123, 2121

\bibitem[\protect\citeauthoryear{Soderblom et al.}{1993}]{Soderblom93} Soderblom D. R., Jones B. F., Balachandran S., Stauffer J. R., Duncan D. K., Fedele S. B., Hudon J. D., 1993, AJ, 106, 1059

\bibitem[\protect\citeauthoryear{Soker \& Kashi}{2011}]{Soker11} Soker N.,  Kashi A., 2011, arXiv:1107.3454

\bibitem[\protect\citeauthoryear{Steffen et al.}{2010}]{Steffen10} Steffen W., Koning N., Wenger S., Morisset C., Magnor M., 2010, arXiv:1003.2012


\bibitem[\protect\citeauthoryear{Townsley \& G\"ansicke}{2009}]{townsley09}
Townsley, D. M., G\"ansicke, B. T., 2009, ApJ, 693, 1007

\bibitem[\protect\citeauthoryear{Tylenda \& Soker}{2006}]{Tylenda06} Tylenda R., Soker N., 2006, A\&A, 451, 223

\bibitem[\protect\citeauthoryear{Tylenda et al.}{2011}]{Tylenda11} Tylenda R., Kami{\'n}ski T., Schmidt M., Kurtev R., Tomov T., 2011, A\&A, 532, 138

\bibitem[\protect\citeauthoryear{Weight et al.}{2003}]{Weight03} Weight A., Evans A., Albinson J. S., Krautter J., 1993, A\&A, 268, 294

	

\end{thebibliography}
\end{document}